\documentclass[12pt,letterpaper]{article}
\usepackage{jheppub}

\usepackage{graphicx} 
\usepackage{subfigure}
\input{epsf}
\usepackage{amsthm}
\theoremstyle{plain}
\newtheorem*{proposal}{Proposal}

\title{Holography for a small world}
\author[]{Vladimir Rosenhaus}
\affiliation[a]{Center for Theoretical Physics and Department of Physics,\\
 University of California, Berkeley, CA 94720, U.S.A.}
\emailAdd{vladr@berkeley.edu}

\begin{document}

\abstract{The holographic principle asserts that the complete description of the interior of a sphere is a theory which not only lives on the surface of the sphere, but also has $A/4$ binary degrees of freedom. In this context we revisit the question of whether AdS/CFT is fully holographic. We construct states which are localized deep in the interior yet are encoded on short scales in the CFT, seemingly in conflict with the UV/IR prescription. We make a proposal to address the more basic question of which CFT states are sufficient to describe physics within a certain region of the bulk.}

\maketitle

\section{Introduction}

The spherical entropy bound \cite{Bek81, Bek94b} gives a remarkable bound on the number of states contained within a sphere of area $A$. The holographic principle~\cite{Tho93,Sus95, Bou99b,RMP} builds on it to make the far more extraordinary assertion that there exists a theory that lives on the sphere and describes all of physics within the sphere, and accomplishes this feat while only having a Hilbert space of dimension $\exp(A/4 l_{pl}^2)$. 

AdS/CFT has only partially realized the holographic principle. The CFT does live on the boundary of AdS, but its Hilbert space is infinite-dimensional. The boundary of the region described has an area which is also infinite, and comparing two infinities is not meaningful. 

\begin{figure}[tbp]
\centering
\subfigure[]{
	\includegraphics[width=2in]{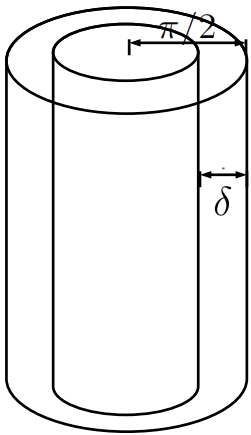}
	}
\caption{The interior of the larger cylinder represents AdS. The vertical direction is time and the radial direction is $\rho$. The CFT lives on the boundary ($\mathcal{R} \times S^{d-1})$ which is at $\rho = \pi/2$. In order for AdS/CFT to fully realize the holographic principle, some theory which has a Hilbert space of dimension $\exp(A/4 l_{pl}^{d-1})$, where $A$ is the area of the sphere at $\rho = \pi/2-\delta$, must be extracted from the CFT. This theory would live on the sphere at $\rho = \pi/2 - \delta$ (inner cylinder) and would need to be able to fully describe the interior: $0 < \rho < \pi/2-\delta$.}  \label{fig:Cylin}
\end{figure}

To confirm the holographic principle in AdS we must extract from the CFT a theory of the appropriate dimension that is capable of fully describing the innermost region of AdS, out to a sphere of area $A$ (see Fig.~\ref{fig:Cylin}). The CFT on a lattice is a natural candidate, and was the basis of the UV/IR proposal of Susskind and Witten~\cite{SusWit98}. However, the UV/IR proposal is known to have some limitations, motivating us to analyze its validity in a range of contexts. 

In the first part of this paper we will construct certain states which explicitly violate UV/IR; the bulk states will be within the sphere of area $A$, yet their boundary image will have features on scales far smaller than the lattice spacing. Our examples will demonstrate UV/IR is violated for both particles which stay inside the sphere and those which enter and leave; for particles that are both relativistic and non-relativistic; for both static states and dynamic states. 

In the second part of the paper we start afresh in looking for the correct theory describing the interior of the sphere of area $A$. We make a proposal for an answer to the question: what CFT states should this theory contain? This is a much easier question than constructing a full theory; a full theory requires having not only the states, but also the observables and a way of doing time evolution. It is, however, a more difficult question than finding just the states confined within the sphere that the Bekenstein bound specializes to. If the interior of the sphere is truly a holographic image, the hologram must be able to describe relativistic particles which enter and leave the sphere. We will discuss how our proposal avoids the difficulties UV/IR encountered. We also comment on the necessity of excluding ultraboosted states from the description.

The paper is organized as follows. In Sec.~\ref{sec1} we review the UV/IR proposal. In Sec.~\ref{sec2} we review the well-known example of a relativistic particle oscillating inside of AdS. The boundary image of the bulk gravitational field induced by the particle is the energy-momentum tensor concentrated to a thin shell, and consequently strongly violating UV/IR. In Sec.~\ref{sec3} we begin our study of scalar field wave packets. We construct a well localized and highly relativistic wave packet which goes inside the bulk IR cutoff, yet on the boundary the expectation value of the CFT operator dual to the scalar field is localized to a region well below the lattice spacing. In Sec.~\ref{sec4} we consider a scalar field with large mass (in AdS units), and consider a mode with angular momentum much larger than the inverse of the lattice spacing but much smaller than the mass. The mode is localized within the central AdS radius, yet the boundary image is on scales below the lattice spacing. In Sec.~\ref{sec5} we consider a general solution of the  Klein-Gordon equation and show that in order for the CFT to not lose information about it when placed on a lattice would require a lattice spacing far smaller than the one prescribed by UV/IR. In Sec.~\ref{sec6} we discuss the possibility that the information contained in local CFT operators which went missing when a UV cutoff was placed is in fact retained in some other ``precursor'' operators. In Sec.~\ref{sec7} we present our proposal for which CFT states are sufficient to fully describe the interior of the sphere of area $A$. 

The question of the validity of the ``scale/radius'' relation is related but somewhat different from the question of the validity of the UV/IR proposal. As a side note, in Sec.~\ref{sec:ScaleRadius} we comment that scale/radius does not follow from the rescaling isometry of the Poincare patch metric, nor is it generally valid. Our example in Sec.~\ref{sec3} explicitly violates scale/radius.

\section{UV/IR}
\subsection{Review of UV/IR} \label{sec1}
In this section we review the UV/IR proposal~\cite{SusWit98}. 

Consider AdS in global coordinates
\begin{equation} \label{eq:rhoCoord}
ds^2 = \frac{L^2}{\cos^2\rho}(-d\tau^2 + d\rho^2 + \sin^2\rho\ d\Omega_{d-1}^2). 
\end{equation}
The CFT lives on the boundary of AdS, at $\rho = \pi/2$, on a sphere that has been conformally rescaled to have radius $1$. The UV/IR prescription seeks to provide a theory that can describe the interior of AdS for all $\rho <\pi/2 - \delta$, where $\delta \ll 1$. The full CFT is of course capable of doing this, however it has an infinite-dimensional Hilbert space and the holographic principle tells us a finite-dimensional Hilbert space should suffice. 

In many computations in AdS/CFT, a bulk quantity that is IR divergent is dual to a CFT quantity that is UV divergent. For instance, the divergence of the length of a string ending on the boundary is dual to the divergent self-energy of a quark. This observation motivated \cite{SusWit98} to propose that the theory we are looking for is the CFT placed on a lattice. Since the CFT is an $SU(N)$ gauge theory, \cite{SusWit98} wanted to count $N^2$ degrees of freedom per lattice site. The lattice spacing is then fixed by having the number of CFT degrees of freedom match the area in Planck units of the sphere at $\rho = \pi/2-\delta$:
\begin{equation}
\frac{1}{l_{pl}^{d-1}} \left(\frac{L}{\delta}\right)^{d-1} .
\end{equation}
Using the relation $N^2 = (L/l_{pl})^{d-1}$, we see that the lattice size is fixed to be $\delta$. 

Since each of these degrees of freedom has, like a harmonic oscillator, an infinite-dimensional Hilbert space, \cite{SusWit98} needed to further impose that each oscillator can only be excited to the first few energy levels. This amounts to imposing an energy density cutoff of  $(1/\delta)^{d}$. The energy density cutoff will not be important for us since UV/IR will face difficulties already at the stage of the spatial lattice. 

We should note that the terminology ``UV/IR'' is used in a range of contexts. For us, UV/IR will mean the specific proposal reviewed above of how to truncate the CFT and still be able to describe the portion of the interior, $0<\rho < \pi/2-\delta$. 

\subsection{Comments on Scale/Radius} \label{sec:ScaleRadius}
In this section we include a few comments on the ``scale/radius'' relation and how it relates to UV/IR. The scale/radius relation is the statement that an object close to the boundary should be dual to a CFT state with small spatial extent, whereas an object deep in the bulk should be dual to a CFT state of large spatial extent. 

The degree to which the scale/radius relation is generally valid is somewhat orthogonal to the one of UV/IR we are interested in. In particular, that a CFT state may have a certain spatial extent does little in terms of telling us on what scale the state has features, and hence what kind of lattice on the CFT would be sufficient to accurately describe this state.

In fact, we see no basis for even scale/radius being valid in any dynamical context. The scale/radius relation is motivated by the isometry of AdS under the rescaling of Poincare coordinates: $(t,x,z) \rightarrow (\lambda t, \lambda x, \lambda z)$. Consider some bulk solution; for instance a solution $\Phi_0(x,t,z)$ of the scalar wave equation. This configuration will have a boundary imprint $\phi_0(x,t)$ which is obtained by extracting the coefficient of the decaying tail of $\Phi_0$ near the boundary. The AdS/CFT dictionary tells us $\phi_0(x,t)$ is equal to $\langle O(x,t) \rangle$ on the CFT. The isometry of AdS means one can construct a one-parameter family of bulk solutions $\Phi_0 (\lambda x, \lambda z, \lambda t)$ and these will have a boundary imprint $\phi_0(\lambda x, \lambda t)$. Thus as $\lambda$ is increased, $\Phi_0$ will be peaked deeper in the bulk and the boundary imprint grows in spatial extent by a factor $\lambda$.

However, for the scale/radius relation to be relevant in explaining the emergent radial direction as an energy scale in the CFT, it would need to be true as a dynamical statement. AdS/CFT is a complete duality; not only can a bulk state be mapped to a boundary state, but also the equivalence must be maintained under time evolution. So it is not sufficient to show that processes characteristically closer to the the boundary have a smaller boundary size at some point in their evolution. One must show that throughout the evolution of a localized bulk object, its radial location is correlated with the spatial extent of the CFT image. The isometry of AdS under rescalings only implies the former and not the later.

Our example in Sec.~\ref{sec2} of an oscillating particle (and the growing/contracting shell to which it is dual) is generally regarded as consistent with scale/radius. However, this is true in a trivial way: the oscillating particle is obtained from the static one at $\rho=0$ through a scale transformation in Poincare coordinates.  Our example in Sec.~\ref{sec3} will explicitly violate scale/radius. 
 
\section{Oscillating Particle} \label{sec2}

\begin{figure}[tbp]
\centering
\subfigure[]{
	\includegraphics[width=2in]{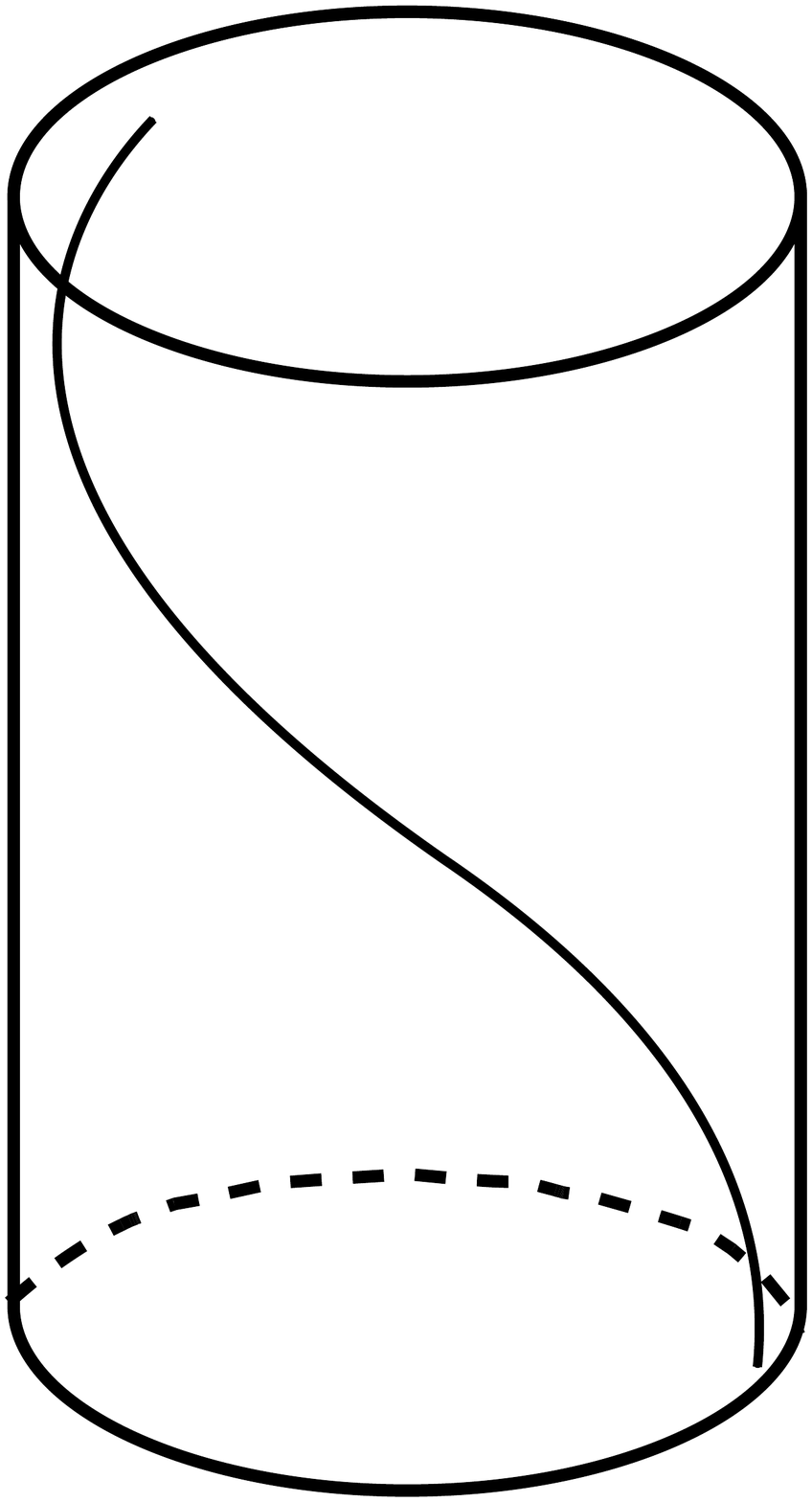}
	}
	\hspace{.2in}
		\subfigure[]{
	\includegraphics[width=2in]{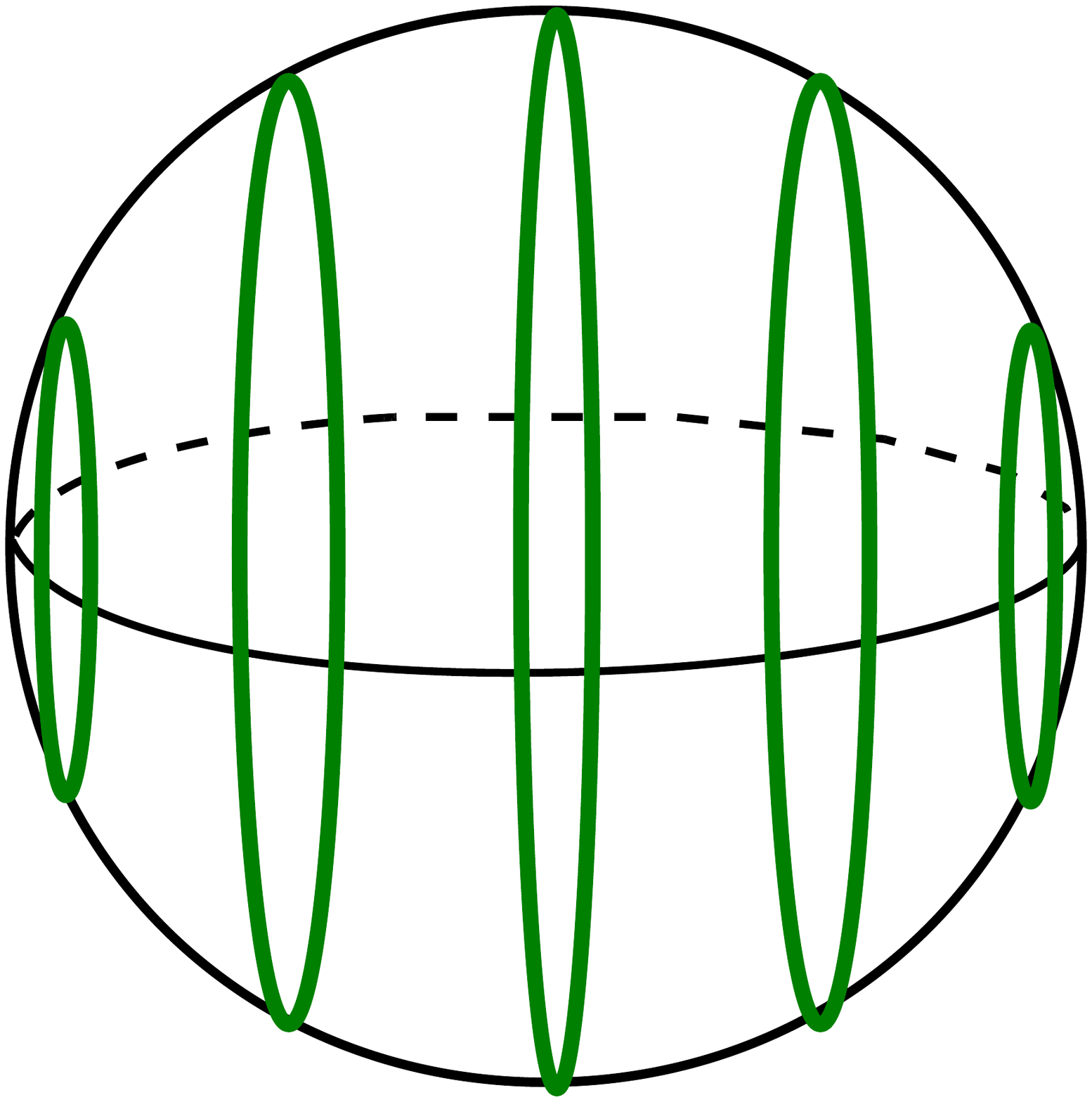}}
\caption{(a)  A relativistic particle following a radial geodesic inside of AdS. Its closest approach to the boundary is $\rho = \pi/2 - m/E$. (b) The sphere shown is the $S^{d-1}$ of the CFT (it is represented as an $S^2$; in (a) we were only able to draw the boundary as an $S^1$). On the CFT the particle is represented by a thin shell of $\langle T_{\mu \nu} \rangle$. This shell is shown at multiple instances of time. As the particle starts near the boundary, the shell is small and near the right pole of the sphere. The shell grows as the particle falls toward smaller $\rho$. As the particle passes through $\rho = 0$ the shell wraps the entire $S^{d-1}$. As the particle moves out again to larger $\rho$, the shell contracts. Crucially, the thickness of the shell is $m/E$.}  \label{fig:Osc}
\end{figure}

In this section we review the example of a relativistic particle oscillating inside of AdS (see Fig. \ref{fig:Osc}). The backreaction of the particle changes the metric and induces a nonzero  $\langle T_{\mu \nu} \rangle$ on the boundary. As the particle passes through the center, $\langle T_{\mu \nu} \rangle$ remains concentrated on a thin shell. It was pointed out in \cite{PolSus99} that this example is in tension with UV/IR.

It will be convenient to consider AdS in coordinates
\begin{equation} \label{eq:globalCoord}
ds^2 = -\left(1+\frac{r^2}{L^2}\right)d\tau^2 +\frac{dr^2}{1+\frac{r^2}{L^2}} + r^2 d\Omega_{d-1}^2.
\end{equation}
These coordinates are related to the (\ref{eq:rhoCoord}) coordinates by $r = L \tan \rho$. The boundary of AdS is at $r \rightarrow \infty$. In this limit the metric (\ref{eq:globalCoord}) asymptotes to 
\begin{equation} \label{eq:limit}
ds^2 = r^2 \left(- \frac{d\tau^2}{L^2} + d\Omega_{d-1}^2\right).
\end{equation}
The metric of the $S^{d-1} \times \mathcal{R}$ on which the CFT lives is obtained by conformally rescaling (\ref{eq:limit}) by a factor of $1/r^2$, giving a sphere of radius equal to $1$. 

A particle of mass $m$ oscillating in AdS (Fig. \ref{fig:Osc}a) satisfies the geodesic equation
\begin{equation}
\dot{r}^2  = \left(\frac{E}{m}\right)^2 -1 -\left(\frac{r}{L}\right)^2 ,
\end{equation}
where $E$ is the energy of the particle with respect to the timelike Killing field,
\begin{equation}
E/m = (1+ r^2/L^2) \dot{\tau} .
\end{equation}
The proper energy of the particle at the center of AdS is equal to $E$, and the CFT energy of this state is $E L$. The largest $r$ the particle reaches is $r_{max} \approx L E/m$, where we have assumed $E \gg m$. In terms of coordinates (\ref{eq:rhoCoord}), $\rho_{max} = \pi/2 - \alpha$, where we defined $\alpha \equiv m/E$. Since the particle is relativistic, $\alpha \ll 1$. 

The computation of  $\langle T_{\mu \nu}\rangle$ for this state was done by Horowitz and Itzhaki \cite{HorItz99}, and we collect their results in Appendix \ref{secAppendix}. In Fig. \ref{fig:Osc}b we have sketched how $\langle T_{\mu \nu}\rangle$ evolves. The energy of the CFT state is concentrated on a shell of thickness $\alpha$. As the particle falls into the bulk the shell expands, reaching a maximum size when the particle reaches $r =0$. The shell then contracts as the particle moves out towards larger $r$. Crucially, the thickness of the shell is equal to $\alpha$. 

UV/IR tells us that the CFT on a lattice with spacing $\delta$ should describe the bulk out to $\rho = \pi/2 - \delta$. Thus, it must describe the oscillating particle which goes through $\rho$ smaller than the cutoff. Yet, if we choose $\alpha \ll \delta$, then the shell has a width far smaller than the lattice spacing. The extent to which the cutoff CFT fails to describe the particle can be made arbitrarily large by making $\alpha$ small. The extreme case would be a massless particle traveling through the bulk. Its boundary dual is a shell that is completely localized on the lightcone $\theta = \tau$.

\section{Scalar Field Solutions}
The example in Sec.~\ref{sec2} of an oscillating particle presents a constraint on imposing any kind of lattice on the CFT. However, this example is a bit special and we would like to have a larger set of examples to test UV/IR. That is what we do in this section. Instead of working with the gravitational field, we will consider a free scalar field $\phi$, 
\begin{equation} \label{eq:Wave}
(\Box + m^2) \phi = 0.
\end{equation}
The CFT operator $O$ is dual to $\phi$. Throughout this section we will consider some solutions of (\ref{eq:Wave}) and look at $\langle O \rangle$ for these states. We will be interested in solutions of (\ref{eq:Wave}) which either at some time, or for all time, are contained within the bulk region $\rho <\pi/2 - \delta$. In our examples we will construct states that do this and also have $\langle O \rangle$ that is concentrated on scales much less than $\delta$. 

\subsection{Relativistic Wave Packet} \label{sec3}

\begin{figure}[tbp] 
\centering
\subfigure[]{
	\includegraphics[width=3.1in]{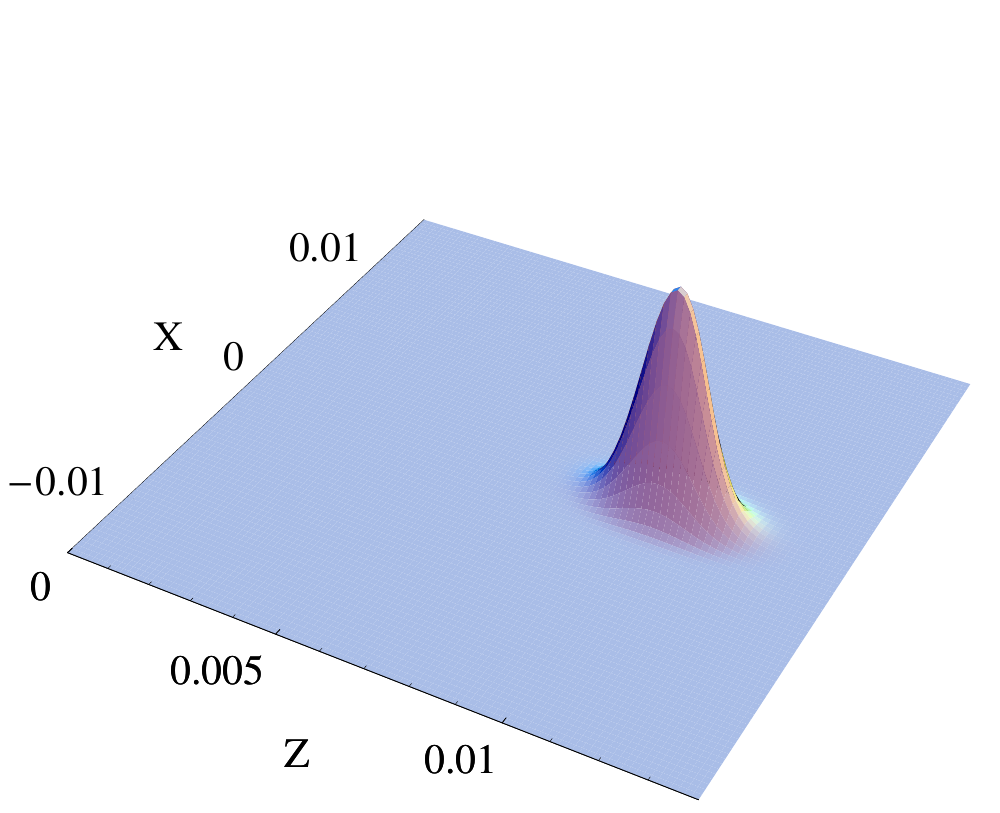}
	}
	\hspace{.2in}
		\subfigure[]{
	\includegraphics[width=2.5in]{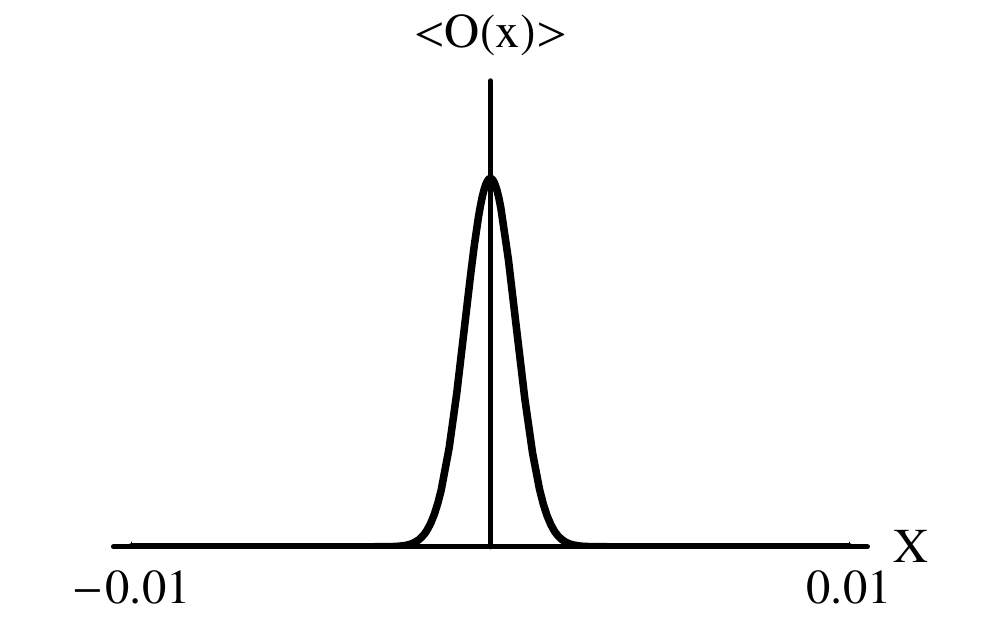}}
\caption{(a) The profile of a localized wave packet (\ref{eq:packet}) traveling away from the boundary ($q_0=10^6,\ \sigma_x = \sigma_z= 10^{-3})$ shown at time  $t=10^{-2}$.  It is composed of modes highly oscillatory in the $z$ direction, of wavenumber peaked around $q_0$. (b) The CFT image $\langle O(x) \rangle$  at this time (given by (\ref{eq:Opacket2})).} \label{fig:packets}
\end{figure}

In this section we would like to construct a wave packet that travels from near the boundary of AdS into the bulk. We would like this packet to remain well concentrated and have negligible spread as it propagates into the bulk. It is familiar from wave physics that packets with large momentum in one direction have, for a long period of time, negligible spreading in the transverse directions. In AdS we can construct packets with a similar property (shown in Fig. \ref{fig:packets}a). For these packets we will want to find how $\langle O \rangle$ evolves with time. Near the boundary ($\rho \approx \pi/2$), the field value will decay as $(\cos \rho)^{\Delta}$. The AdS/CFT dictionary tells us that the coefficient in front of this term determines $\langle O \rangle$. So to find $\langle O \rangle$ we need to find the behavior of the tail of the wave packet at $\rho \approx \pi/2$. It should be pointed out that just because the peak of the packet in the bulk has negligible transverse spread does not yet imply the tail will also have negligible spread with time; although we will see this is what occurs in AdS. 

The packets we will construct can be used to violate UV/IR by arbitrarily large amounts. For some given lattice spacing $\delta$, we make a packet with transverse spread much less than $\delta$. We then make it sufficiently energetic so that it doesn't spread for a long time. On the CFT, $\langle O \rangle$ remains concentrated to a region much less than $\delta$, even when the packet is at $\rho < \pi/2 - \delta$. 

To construct the packets it will be convenient to use Poincare coordinates, which are a good approximation to global coordinates near the boundary and for small angular spread. Inserting $\rho = \pi/2 - z$ into (\ref{eq:rhoCoord}) gives
\begin{equation} \label{eq:Poincare}
ds^2 = \frac{L^2}{z^2}\left(-dt^2+ dz^2 + dx^2\right),
\end{equation} 
where the angular coordinate is now $x$ and the time coordinate has been relabeled $t = \tau$. 
In these coordinates the mode solutions to (\ref{eq:Wave}) are 
\begin{equation} \label{eq:PoinModes}
\varphi_{q k}(x,t,z) = z^{d/2} J_{\nu}(q z) e^{i k x - i \omega_{q k} t} ,
\end{equation}
where $ 0<q<\infty$, the energy  is $\omega_{q k}^2 = q^2 + k^2$, and $\nu = \Delta - d/2$. The conformal dimension $\Delta$ is taken to be of order $1$, and is related to the mass $m$ through $ \Delta (\Delta - d) =m^2 L^2$. 

We now construct our packet out of the modes (\ref{eq:Poincare}) peaked around a large $q=q_0$ with spread in $q$ of $\sigma_z^{-1}$, and peaked around $k=0$ with spread in $k$ of $\sigma_x^{-1}$:
\begin{equation} \label{eq:packet}
\Phi (x,t,z) = \int{dq \ dk \sqrt{q} \ \varphi_{q k} (t,x,z) \ e^{-k^2 \sigma_x^2/4}\ e^{- (q-q_0)^2 \sigma_z^2/4}},
\end{equation}
where $q_0 \gg \sigma_z^{-1}, \sigma_x^{-1} \gg 1$. To verify the packet has the behavior we desire, we make some simplifications to allow us to evaluate the integrals in (\ref{eq:packet}). First we notice that particles in AdS only behave  differently from those in Minkowski space when they get close enough to the boundary such that the gravitational potential energy becomes comparable to their kinetic energy. In terms of the modes (\ref{eq:PoinModes}) this is reflected in the the $z$ component: 
\begin{equation}
J_{\nu}(qz) \sim \frac{1}{\sqrt{qz}} e^{i q z} , \ \ \  \text{for} \ \  qz \gg \nu ~.
\end{equation}
This means we can think of $q$ as a kind of radial momentum. From the form of (\ref{eq:packet}) we see that $q$ is peaked around $q_0$ with spread $\sigma_z^{-1}$ which is much less than $q_0$. Thus for $z \gg \nu/q_0$ we can approximate (\ref{eq:packet}) by 
\begin{equation} \label{eq:packet2}
\Phi (x,t,z) \approx z^{(d-1)/2} \int{dq\ dk\ e^{i q z}\ e^{-k^2 \sigma_x^2/4}\ e^{i k x}\ e^{- (q-q_0)^2 \sigma_z^2/4}\ e^{- i \omega_{q k} t}}.
\end{equation}
Aside from the uninteresting power of $z$ in front, (\ref{eq:packet2}) is of the same form as a packet propagating in the $z$ direction in Minkowski space. Since in (\ref{eq:packet2}), $k \lesssim \sigma_x^{-1} \ll q_0$, we approximate
\begin{equation} \label{eq:omegaqk}
\omega_{qk} = \sqrt{q^2+k^2} \approx q + \frac{k^2}{2 q_0}.
\end{equation}
This allows us to separate the $q$ and $k$ integrals in (\ref{eq:packet2}), and easily evaluate the integral over $q$,
\begin{equation}
\Phi (x,t,z) = z^{(d-1)/2}\ \psi(x,t)\ e^{-(z-t)^2/\sigma_z^2}\ e^{i q_0 (z-t)},
\end{equation}
where 
\begin{equation} \label{eq:psi}
\psi(x,t) = \int{dk\  e^{i k x}\  e^{-k^2 \sigma_x^2/4}\  e^{-i  \frac{k^2 t}{2 q_0}}}.
\end{equation}
We have dropped constants and have labeled the transverse spread as $\psi$ because, as can be seen from the energy (\ref{eq:omegaqk}), it satisfies the non-relativistic Schrodinger equation for a particle of mass $q_0$. Evaluating (\ref{eq:psi}) thus gives the familiar answer, 
\begin{equation}
\psi(x,t) = \frac{1}{\sigma_x \sqrt{1+\frac{2 i t}{ \sigma_x^2 q_0}}} \exp\left(- \frac{x^2}{\sigma_x^2 (1+ \frac{2 i t}{\sigma_x^2 q_0})}\right),
\end{equation}
showing the packet's spread with time is the expected, 
\begin{equation}
\sigma_x(t) = \sigma_x \sqrt{1+\left(\frac{2 t}{ q_0 \sigma_x^2}\right)^2}.                                                               
\end{equation}
This shows that the transverse spread is negligible for times $t \lesssim q_0 \sigma_x^2$. 

We would now like to evaluate $\langle O(x,t)\rangle$. Noting that for small $z$ the Bessel function can be approximated as 
\begin{equation}
J_{\nu}(qz) \sim (qz)^{\nu}  \ \ \  \text{for} \ \ q z< \nu \ ,
\end{equation}
we obtain $\langle O(x,t)\rangle$ from the $z\rightarrow 0$ limit of $z^{-\Delta} \Phi(x,t,z)$. Using (\ref{eq:packet}) we find,

\begin{equation} \label{eq:Opacket}
\langle O(x,t) \rangle = \int{dq\ dk \ q^{\nu+1/2}\ e^{-k^2 \sigma_x^2/4 }\ e^{i k x} \ e^{-(q-q_0)^2 \sigma_z^2/4}\ e^{- i \omega_{q k}t}\ } .
\end{equation}
We now perform the same simplifications we used on (\ref{eq:packet}), separating the integrals over $q$ and $k$, and finding
\begin{equation} \label{eq:Opacket2}
\langle O(x,t) \rangle = e^{- i q_0 t} \psi(x,t),
\end{equation}
where $\psi(x,t)$ is given by the same expression as (\ref{eq:psi}) before, and we have ignored terms that are independent of $x,t$. Eq. \ref{eq:Opacket2} shows that $\langle O(x,t) \rangle$ remains localized within $|x|\lesssim \sigma_x$ for a time $t \lesssim q_0 \sigma_x^2$,  as well as being highly oscillatory with time.\footnote{Since our interest is the CFT defined on the sphere of the global AdS boundary, our equations should only be used for small times, $t\ll 1$, where the approximation of the Poincare patch metric is valid. To follow the packet deeper into the bulk we would need to use the true global AdS evolution. Since it will take the packet a time of $\pi/2$ to reach the center of AdS, we expect its transverse spread even at the center to remain comparable to $\sigma_x$ (with a sufficiently large choice of $q_0$). It is interesting to consider what would happen if we were actually interested in the CFT theory on the Minkowski boundary of the Poincare patch. Then as the packet moves towards the Poincare horizon located at infinite $z$, the CFT profile would spread over the entire $x$ axis. This is not surprising; in this case the packet is traveling for an infinite amount of Poincare time.}
A wave packet and its dual $\langle O\rangle$ at an instant of time are shown in Fig. \ref{fig:packets}. We see that the UV/IR prescription can be violated by arbitrarily large amounts with this example. To decrease the boundary size, $\sigma_x$, we simply need to increase $q_0$ in order to maintain $q_0 \gg \sigma_x^{-1}$. 
\enlargethispage{.35cm}

Since we are taking $q_0$ to be large, one might worry that if there is other matter around then gravitational effects might invalidate the use of the free wave equation (\ref{eq:Wave}). However, any such effects would occur on scales set by the Planck scale. The Planck scale is related to the AdS scale by a power of $1/N$.
Crucially, the lattice size on the boundary that UV/IR prescribes is independent of $N$, allowing us to take $N$ as large as we want. 

\subsection{Non-relativistic Mode} \label{sec4}

In Sec.~\ref{sec2} and Sec.~\ref{sec3} we considered relativistic particles with large momentum in the radial direction. In this section we go to the other extreme of a non-relativistic particle with angular momentum. The radial location of the particle is determined by two competing forces. AdS is confining, and the more massive the particle, the more it is confined to the center of AdS. On the other hand, the centrifugal barrier from the particle's  angular momentum pushes it out to larger radius. We will see that in order for the boundary image to have features on small scales, the particle needs to be given large angular momentum. By dialing the particle's mass to be sufficiently large, we  make the particle well-confined to the center of AdS, and hence violate UV/IR.

\begin{figure}[tbp]
\centering
	\includegraphics[width=3in]{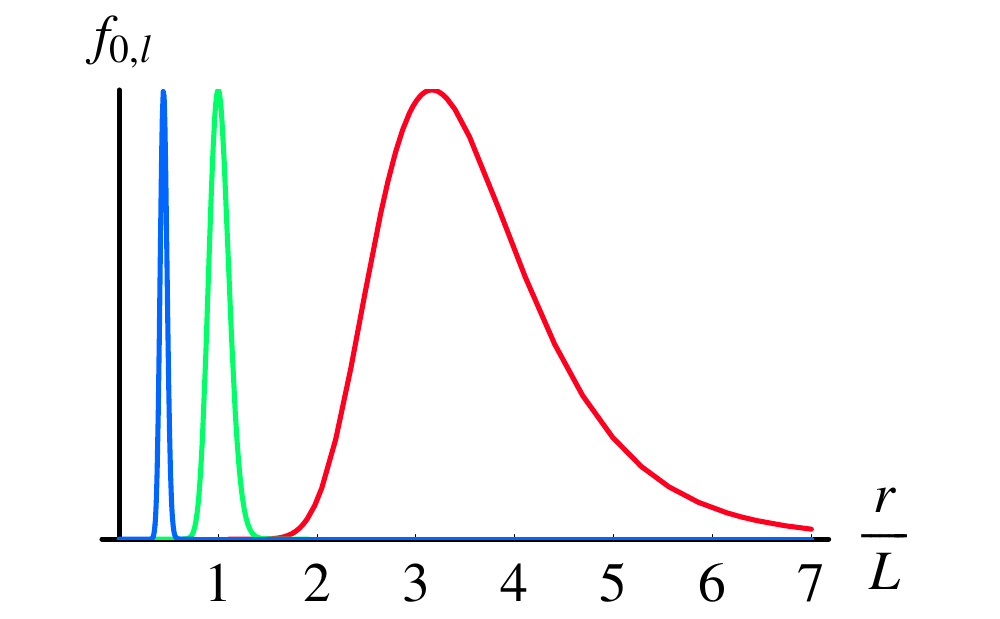}

\caption{A plot of the radial modes $f_{0 l}(\rho)$ in (\ref{eq:Modenl}) for fixed $l$ ($l=100$) and several different choices of $\Delta$ ($\Delta = 10,\ 100,\ 500$, in colors red, green, blue, respectively). The variable on the horizontal axis is the radial coordinate $r/L = \tan \rho$. The plot shows that increasing $\Delta$ leads to stronger confinement to the center of AdS. To violate UV/IR we choose a mode with $\Delta \gg l\gg 1/\delta$.}  \label{fig:MassModes}
\end{figure}

As in the previous section, we consider a scalar field and for simplicity our equations are in $AdS_3$. The mode solutions of the Klein-Gordon equation (\ref{eq:Wave}) in global coordinates (\ref{eq:rhoCoord}) are given by, 
\begin{equation} \label{eq:Modenl}
\varphi_{n l } ( \rho, \tau, \theta) = f_{n l}(\rho) e^{i l \theta} e^{-i \omega_{n l} \tau} .
\end{equation}
The energy $\omega_{n l} = \Delta + 2n +l$, where $\Delta \approx m L$ for large mass, and the radial part $f_{n l}(\rho)$ is some hypergeometric function whose explicit form we have suppressed. For small $\rho$ AdS looks like Minkowski space in spherical coordinates, and the radial modes reduce to the familiar spherical Bessel functions. 

Consider a bulk solution that consists of just a single mode, (\ref{eq:Modenl}). On the boundary, the expectation value of $O$ for this state is simply proportional to (\ref{eq:Modenl}) with the radial portion removed, 
\begin{equation}
\langle O(\theta, \tau) \rangle \propto e^{i l \theta} e^{-i \omega_{n l} \tau} .
\end{equation}
Let us now put the CFT on a lattice of size $\delta$. Since the CFT lives on a circle of radius $1$, this amounts to restricting to modes with $l \lesssim \delta^{-1}$. Thus, to violate UV/IR we simply need to consider a bulk solution $\varphi_{0 l}$ with $l\gg \delta^{-1}$ and  take the mass of the field sufficiently large to have this mode confined near the center. In other words, $ \Delta \gg l \gg \delta^{-1}$. 
Since this example consists of an energy eigenstate, it illustrates that  UV/IR  is insufficient even for describing static bulk physics which remains confined within $\rho <\pi/2 - \delta$.

\subsection{General Solution} \label{sec5}
We have seen examples of both relativistic and non-relativistic wave packets that cause difficulties for UV/IR. But perhaps these examples are special and for a generic bulk field configuration, the UV regulated CFT will be able to describe it fairly well. This is the question we address in this section. 

Consider some solution $\Phi(\rho, \tau, \theta)$ of the wave equation (\ref{eq:Wave}). Expanding $\Phi$ in terms of modes (\ref{eq:Modenl}), 
\begin{equation} \label{eq:ModeExp}
\Phi(\rho, \tau, \theta) = \sum_{n, l}{c_{n l}\ f_{n l}(\rho)\ e^{i l \theta - i \omega_{n l} \tau}}.  
\end{equation}
Assuming the modes are normalized with respect to the standard bulk Klein-Gordon norm, near the boundary $f_{n l}(\rho)$ takes the form $d_{n l} (\cos \rho)^{\Delta}$, where
\begin{equation} \label{eq:dnl}
d_{n l} = (-1)^n \sqrt{\frac{\Gamma(\Delta + n +|l|)\Gamma(\Delta + n)}{n! \Gamma(\Delta)^2 \Gamma(n+|l|+1)}}.
\end{equation}
We can now find the boundary imprint of this solution,
\begin{equation}\label{eq:O}
\langle O(\tau, \theta) \rangle = \sum_{n, l}{c_{n l}\ d_{n l}\ e^{i l \theta - i \omega_{n l} \tau}}.
\end{equation}

The question now is: for a solution $\Phi$ relevant to the region $\rho < \pi/2 - \delta$, are the important features of the corresponding $\langle O \rangle$ on scales larger than $\delta$? On the boundary, a spatial lattice of size $\delta$ is equivalent to cutoff on $l$  of $\delta^{-1}$. In order to be able to truncate the sum in (\ref{eq:O}) to $l < \delta^{-1}$, it would need to be the case that $c_{nl}$ is suppressed at these large $l$. However, this is clearly not generally true.  At any fixed $\rho$ in the bulk we need a complete basis of modes in the angular direction, which requires retaining modes of arbitrarily high $l$. 

To be more precise, since we don't expect local field theory to be valid in the bulk below the Planck scale, the highest $l$ we would actually need in order to describe $r<R$ would be $R/l_{pl}$. However, this would require a lattice spacing on the CFT that is far smaller than in the UV/IR prescription: in AdS$_{d+1}$/CFT$_{d}$, the lattice we would need to describe $\rho< \pi/2- \delta$ would need to have a spacing of $\delta/N^{2/(d-1)}$.

\section{Precursors} \label{sec6}

We have seen examples of states for which the CFT would lose some of the information contained in $\langle O \rangle$ if it were placed on a lattice. However, perhaps all the information about the state is still present in the CFT on a lattice, but instead of being contained in $\langle O \rangle$ is contained in the expectation value of some other operators? Without a better understanding of the AdS/CFT dictionary this remains a possibility, but one that is difficult to test. In this section we will just make two comments. The first is that we see no compelling reason why upon placing the CFT on a lattice, $\langle O \rangle$ would lose information while these other operators retain it. The second is that if this possibility is realized, then the UV/IR regulated version of AdS/CFT may not be very useful since the aspects of the bulk-boundary dictionary that we knew in the full AdS/CFT would not be applicable to the regulated version.

To discuss how the CFT encodes the bulk it is useful to ask the following question: how would a collection of CFT observers reconstruct some bulk field configuration $\Phi(\rho, \tau, \theta)$? Expanding $\Phi$ in terms of modes $\varphi_{nl}$ as in (\ref{eq:ModeExp}),
the goal of the CFT observers would be to determine all the coefficients $c_{n l}$. Measuring the expectation value of $O$ over the whole sphere at one instant of time, $\langle O(\theta, \tau_0)\rangle$, would give them some of the information, but not all of it. In particular, since the boundary imprint in $\langle O \rangle$ of a single mode is proportional to $\exp( i l \theta - i \omega_{n l} \tau_0)$, the CFT observers would be able to distinguish among the different $l$ quantum numbers. However, all the modes $\varphi_{nl}$ of a fixed $l$ but differing $n$ would give the same imprint in $\langle O(\theta, \tau_0)\rangle$. By measuring $\langle O(\theta, \tau)\rangle$ over a range of times, the CFT observers could start distinguishing the modes with different $n$. Having $\langle O(\theta, \tau)\rangle$ for all $\theta$ and a time of $\Delta \tau = \pi$ would allow for a full reconstruction of $\Phi$. 

If AdS/CFT is complete then there should be a faster way of reconstructing the bulk; there should exist some CFT operators, named ``precursors'' in  \cite{PolSus99}, which can be measured at one instant of time and immediately fully determine  $\Phi$. The precursor operators are expected to be highly nonlocal and their actual form has remained a mystery. In certain dynamical contexts precursors are essential. For instance, if two wave packets collide in the center of AdS, it takes a time of $\pi/2$ before causality allows the result of the collision to propagate to the boundary and become encoded in $\langle O \rangle$. During that $\pi/2$ interval of time, the result of the collision is encoded exclusively in the precursors. 

These considerations lead us to believe that if we were to consider the CFT on some subset of the boundary, $\mathcal{R}\times S^{d-1}$, that does not contain the full time direction, then the only way to reconstruct some portions of the bulk would be by measuring precursors. However, this is not the situation UV/IR presents: the UV regulated CFT in UV/IR still lives on the full time direction.\footnote{In UV/IR an energy density cutoff is prescribed for the CFT in addition to the spatial lattice. The energy cutoff can be regarded as making the time direction into a lattice. Thus, if in our examples UV/IR had failed due to the energy density cutoff, then the need for precursors might have been a good explanation. However,  UV/IR failed already at the stage of imposing the spatial lattice.} Thus it would be unclear why precursors are essential in the UV/IR regulated version of AdS/CFT in contexts where they were not needed in the full AdS/CFT.

Nevertheless, it may be that precursors save UV/IR. In this case, we should ask what the bulk-boundary dictionary would be for the UV/IR regulated AdS/CFT. In the full AdS/CFT one can construct a mapping, perturbatively in $1/N$, between local bulk operators $\hat{\phi}(\rho, \tau, \theta)$ and the CFT operator $O$ smeared over space and time with some kernel. To leading order in $1/N$, this takes the form
\begin{equation} \label{eq:Kabat}
\hat{\phi}(\rho, \tau, \theta) = \int{ d \tau' d \theta'\ K(\rho, \tau, \theta| \tau ', \theta')\ O(\tau', \theta')} .
\end{equation}
The actual form of $K$ was worked out in \cite{HamKab05, HamKab06}, where the authors also pointed out that a UV cutoff on the CFT would destroy the mapping (\ref{eq:Kabat}). Namely, \cite{HamKab05} showed that the divergence in the two-point function of $\hat{\phi}$ at coincident points can only occur from the UV divergence of the two-point function of $O$. Thus, if the theory dual to the inner portion of AdS is the CFT with a UV cutoff, then the only way to express the bulk operators $\hat{\phi}$ in terms of CFT operators is directly in terms of the precursors.

\section{A Proposal for Finding the States} \label{sec7}
We are interested in finding a theory which is capable of describing all the physics that can occur in the interior of a sphere of area $A$ in AdS, while having a Hilbert space of the appropriate dimension. The proposal the UV/IR prescription gave is that  the theory describing the interior $\rho < \pi/2-\delta$ is the CFT on a lattice of size $\delta$. We have seen in previous sections that this proposal faces difficulties. In this section we initiate a new search for the desired theory. 

Finding such a theory is a difficult problem, and we will only try to address a more elementary question: what are the CFT states that this theory would contain? In other words, what is the set of CFT states that are sufficient to cover everything that could possibly happen in the region $0<\rho<\rho_0$? 

We need to know what CFT states are dual to a chair in the center of AdS, or a star, or any object contained within $\rho<\rho_0$. But we also need much more than that. The sphere at $\rho=\rho_0$ that we are discussing is imaginary; we aren't literally putting a shell there. So a particle can fly in and out of the region $\rho<\rho_0$, or two particles can come in, collide, and leave. We must include the CFT states dual to these processes. This is an essential point: if this ``small world'' consisting of $\rho<\rho_0$ is truly a holographic image, every physical process which occurs inside should be encoded in the hologram. The difficult part is that we should choose carefully which CFT states we are keeping, since the holographic principle only allows us to include $\exp(A/4l_{pl}^{d-1})$ independent states. We should note that there is no guarantee the question being posed has an answer. It may just be that one really needs more states, even an infinite number, and there is no hologram for a small world.

We begin this section by reviewing the Bekenstein bound. We then make our proposal for what the CFT states are that we want, and discuss how it passes the tests of the previous sections that UV/IR struggled with. We then discuss a puzzling aspect of our proposal, and indeed any proposal which keeps a finite number of states: ultraboosted states which pass through $\rho<\rho_0$ must be excluded from the description.

\subsection{Proposal}
Consider some weakly gravitating object that has energy $E$ and can be enclosed in a sphere of radius $R$.\footnote{In this paragraph we are in $4$-dimensional Minkowski space.} For instance, the object could be a planet, or a gas inside of a spherical cavity with reflecting walls. The Bekenstein bound gives a bound on the entropy of the object, 
\begin{equation}\label{eq:Bek}
S \leq 2\pi E R ~.
\end{equation}
The requirement that the object not be within its Schwarzschild radius requires $E< R/2 l_{pl}^2$, and transforms (\ref{eq:Bek}) into 
\begin{equation}\label{eq:Bek2}
S \leq \frac{A}{4 l_{pl}^2} ~.
\end{equation}
Eq. (\ref{eq:Bek2}) can be interpreted as saying that if we have a spherical box with enclosing walls, then specifying $A/4 l_{pl}^2$ numbers is sufficient to completely specify what is inside the box. 

We now return to AdS, and consider placing one of these bound systems inside of AdS. Since AdS is confining, placing the system anywhere except the center will incur an energy cost. If we do place it in the center then in order for it to not undergo gravitational collapse, its energy must be bound by its size. These considerations motivate our  proposal for the answer to the question of what CFT states are in the holographic theory. In coordinates (\ref{eq:rhoCoord}), with the AdS radius $L$, and with the CFT living on a sphere of radius $1$,

\begin{proposal}\label{propS}
The CFT states sufficient to fully describe the bulk for all $\rho<\rho_0$  are those with CFT energy less than $M L$, where $M$ is the mass of a black hole of radius $\rho_0$.
\end{proposal}

Let us discuss a few aspects of our proposal. First, we would like to make sure it covers all the states we considered in previous sections. The non-relativistic particle with angular momentum (Sec.~\ref{sec4}) stays confined within $\rho< \rho_0$ and so is the kind of bound state the Bekenstein bound includes, and is easily covered. In Sec.~\ref{sec2} and Sec.~\ref{sec3} we considered a relativistic particle which enters and leaves our sphere. UV/IR had difficulty with it even in the case the particle's energy is measured in AdS units, regardless of how small $l_{pl}$ is. In our proposal, the energy cutoff is in Planck units, and so these relativistic particles are included. 

We also need to verify that our proposal has the correct number of states prescribed by the holographic principle. This follows from consistency of AdS/CFT. On the CFT side we want to count the number of states with energy less than $M L$. This will be dominated by states with energy close to $M L$, and up to order $1$ factors, can be calculated from the standard thermodynamics of a free gas of $N^2$ species.  For $CFT_4$ for large $M$, the $\log$ of the number of such states is of order $N^{1/2}(ML)^{3/4}$. 
We need to compare this with the area in Planck units of a large $AdS_5$ black hole of mass $M$. This is given by 
$(M l_{pl}^3 L^2)^{3/4}/l_{pl}^3$. Using the relation $N^2 = (L/l_{pl})^3$, we see they are equal.\footnote{Although our proposal should be applicable even for small spheres with sub-AdS radius, it is not particularly useful in this case since the number of CFT states is larger than what we want. In particular, there will be states which are dual to graviton energy eigenstates that are delocalized over the entire central AdS region.}

We note that our proposal is more concrete than the heuristic statement that CFT states of low energy are associated with bulk regions of small $\rho$ and states of high energy are associated with the near boundary region, $\rho \approx \pi/2$. In particular, we are not proposing that CFT states of high energy are sufficient to describe the portion of the bulk that is near the boundary. This would obviously be wrong; a relativistic particle oscillating in AdS, like the one in Sec.~\ref{sec2}, transverses nearly all of AdS while having constant energy. All we are saying is that the low energy states may be sufficient to completely describe the small $\rho$ region. 

The hope would be that the meaning of ``fully describe'' in the proposal \ref{propS} would be just that: any bulk observable within $\rho < \rho_0$ could be expressed in terms of some boundary theory containing only the CFT states with energy below the cutoff specified in the proposal. For instance, in the full AdS/CFT, one computes a bulk $n-$point function $\langle \phi(B_1) \phi(B_2) ... \phi(B_n)\rangle$ in terms of a smeared CFT $n-$point function through use of the dictionary (\ref{eq:Kabat}). In our context, we would think of the bulk $n-$point function as the overlap between an $m$ particle state and an $n-m$ particle state. The $m$ particle or $n-m$ particle bulk state can be expressed as a CFT state through use of (\ref{eq:Kabat}). Thus, the bulk $n-$point function is expressed in terms of an overlap between CFT states. For low point functions, our proposal ensures that these CFT states are retained. It is important to note the difference here with UV/IR. If one were to use UV/IR, one would be forced to modify the relation (\ref{eq:Kabat}) and as we saw, one would be unable to correctly reproduce bulk correlation functions. Instead, we correctly reproduce bulk correlation functions by never giving an operator mapping of the form (\ref{eq:Kabat}) that would be valid for the small world hologram, but instead only working with the mapping between bulk and boundary states.

Another distinction between our proposal and UV/IR is that we are taking the low energy states of the CFT, which is a subspace of the CFT Hilbert space. UV/IR, on the other hand, restricted to the long wavelength modes of the CFT - a tensor factor of the Hilbert space. To see the subspace restriction, it is instructive to consider the following example. Consider a state which in the bulk consists of some object (like a planet) localized in the center of a AdS, along with a moon that is orbiting the planet with a large radius. The energy of this state is very large if we take the moon to be far from the planet and close to the boundary. Thus, this state would be excluded from our restricted set of states in (\ref{propS}), which would seem to indicate a difficulty since there is a planet in the center which we are failing to describe. The point, however, is that there exists another state which has just the planet without the moon. This state is low energy, and hence we have retained it in our set of CFT states. In other words, for a bulk state of interest which has something within $\rho<\rho_0$ that we wish to describe, there are many states that differ for $\rho>\rho_0$ but that look approximately the same for $\rho<\rho_0$.

\subsection{Ultraboosted states}

\begin{figure}[tbp] 
\centering
	\includegraphics[width=2.5in]{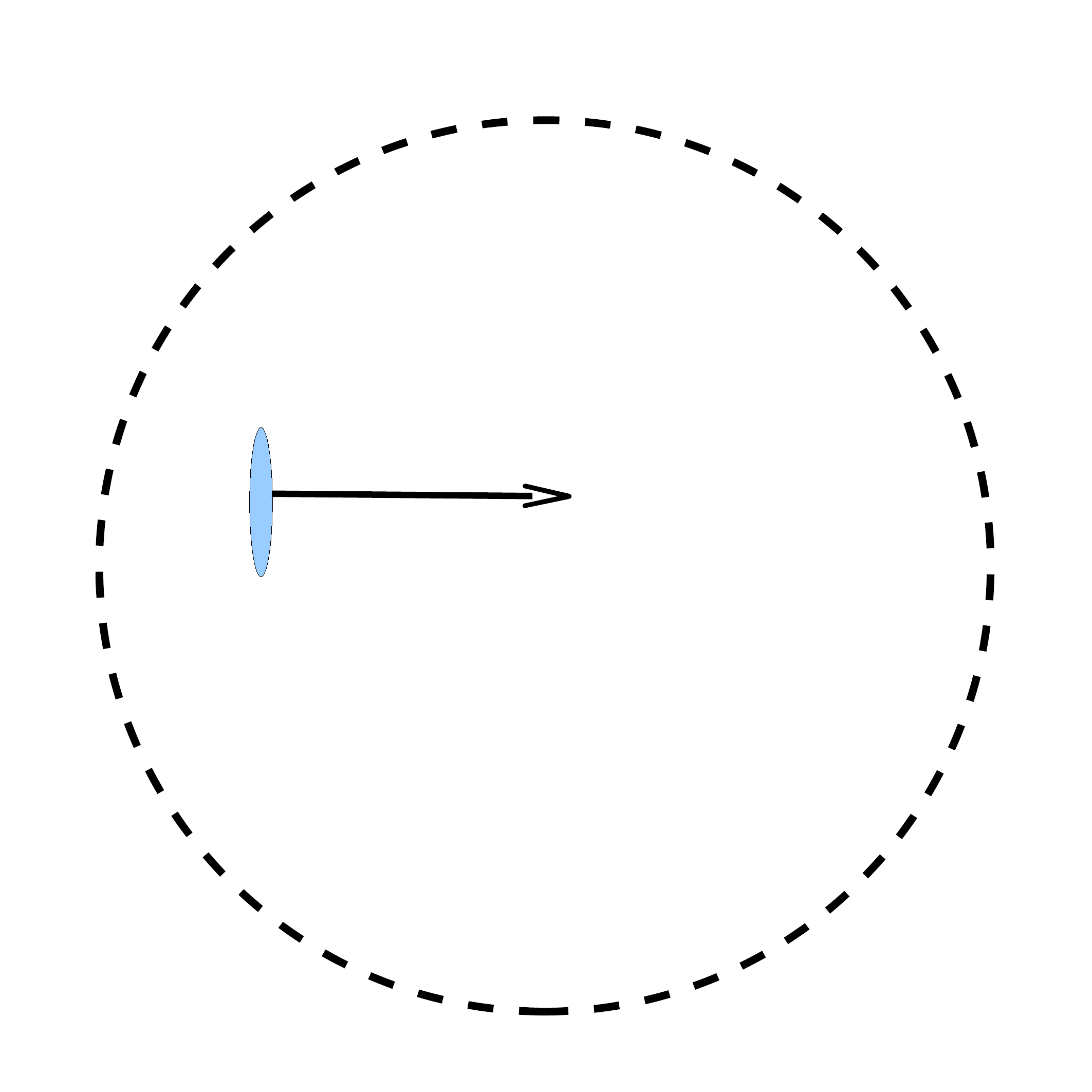}
\caption{A highly boosted state for which the center of mass energy, and hence backreaction, is small. In order for the theory on the sphere to only have a finite number of states, we must exclude this state. A plausible reason is that any attempt to probe it would form a black hole that is much larger than the sphere. } \label{fig:SphereState}
\end{figure}

Having discussed the successes of our proposal, in this section we push it to the brink of failure. Namely, our proposal excludes the high energy states (those with energy greater than $M L$), and we need to know if this is valid. It is certainly justified to exclude a state if it has large backreaction, and this is the first avenue we pursue. Our discussion will focus on the two different frames of reference that are relevant: that of the sphere on which the theory lives, and the center of mass frame.

Consider the state inside the sphere at some instant of time as an object is passing through. The energy, $E$, with respect to the reference frame of the sphere is the energy that appears in our proposal as the cutoff energy determining which states are kept. There is also the center of mass reference frame, and we will let $E_{CM}$ denote the energy in this frame. The backreaction, as characterized by curvature invariants, can be computed in any frame of reference and is easiest to analyze in the center of mass frame.

There are therefore two different possibilities: (a) $E$ and $E_{CM}$ are comparable. (b) $E\gg E_{CM}$. In case (a) our proposal does well, since the states of large energy also have large center of mass energy $E_{CM}$ and consequently collapse into a black hole larger than our sphere. On the other hand, case (b) (see Fig.~\ref{fig:SphereState}) is more interesting. By taking some system of small energy and boosting it by an enormous amount, its energy $E$ can be made arbitrarily large. By going to sufficiently large boost, we can make states which the theory living on the sphere can no longer describe. 

Although it appears odd that sufficiently energetic states are invisible to the theory on the sphere, it is perhaps reasonable. The observables of the theory should correspond to some kind of physical probe of the interior. For instance, the sphere theory could send in particles to probe the interior. These probes will naturally be of low energy, as measured in the sphere reference frame. Now consider the state of an ultraboosted object passing through the interior. Any attempt to probe it in this way will form a large black hole, of radius much larger than where the sphere theory was living. So, in a  physical sense, there is no way for the hologram to be able to describe these states. 

\section{Conclusion}
Our goal has been to see the extent to which UV/IR gives us a theory, with $A/{4 l_{pl}^{d-1}}$ degrees of freedom, that can fully describe the central AdS region out to a sphere of area $A$. We presented some examples which are in tension with UV/IR. We therefore took a step back, turning to the more basic question of which CFT states are sufficient to describe the interior of the sphere. The states of lowest energy seem like a promising candidate, which passes the tests UV/IR struggled with.  Interestingly, an ultraboosted particle passing through would be invisible to the theory on the sphere. 

One may wonder: is there any reason a region of spacetime should be described by a holographic theory with so few degrees of freedom? The most straightforward interpretation of entropy bounds would be that they do nothing more than quantify the simple observation that information requires energy, and too much energy confined to too small of a region leads to gravitational collapse. From this perspective, the holographic principle is an unbelievable extrapolation. And yet, AdS/CFT has partially realized the holographic principle; so perhaps it fully realizes it. Or, perhaps it doesn't and there only exist holograms with infinite information content.

\acknowledgments
I thank R.~Bousso, S.~Leichenauer, and A.~Ovcharova for many helpful discussions. I am grateful to R.~Bousso for years of discussions, advice, and encouragement. I am supported by the Berkeley CTP. 

\appendix

\section{Equations for Energy Shell} \label{secAppendix}

In this appendix we collect several equations from  \cite{HorItz99} relevant for finding $\langle T_{\mu \nu}\rangle$ for the oscillating particle in Sec. \ref{sec2}. The result was already sketched in Fig. $\ref{fig:Osc}$ b.
 
To find $\langle T_{\mu \nu}\rangle$ we need to first find the gravitational field caused by the oscillating particle. The tail of this field at large $r$ will then be proportional to $\langle T_{\mu \nu} \rangle$. Horowitz and Itzhaki \cite{HorItz99} did this calculation in a more elegant way. They first found $\langle T_{\mu \nu} \rangle$ for a stationary particle at $r=0$ and then applied a boost. Their answer is presented in Poincare patch coordinates, since in these coordinates a boost is just a dilatation and easy to implement. The Poincare patch metric is 
\begin{equation} \label{eq:Poin}
ds^2 = \frac{L^2}{z^2}\left(-dt^2+ dx^2 + dz^2\right).
\end{equation} 
Defining $u = t- \sqrt{x^2}$ and $v = t+ \sqrt{x^2}$, the result of \cite{HorItz99} for $AdS_5/CFT_4$ is
\begin{eqnarray}\label{eq:TPoin}
\langle T_{u u} \rangle &=& \frac{8 m \alpha^4}{3 \pi^2} \frac{1}{(\alpha^2 + u^2)^3 (\alpha^2 +v^2)}\\
\langle T_{v v} \rangle &=& \frac{8 m \alpha^4}{3 \pi^2} \frac{1}{(\alpha^2 + v^2)^3 (\alpha^2 +u^2)}\\
\langle T_{u v} \rangle &=& \frac{4 m \alpha^4}{3 \pi^2} \frac{1}{(\alpha^2 + u^2)^2 (\alpha^2 +v^2)^2} .
\end{eqnarray}
In Poincare patch coordinates, the particle starts at $z=\alpha$ at $t=0$ and then falls toward larger $z$. For fixed $v$, $\langle T_{\mu \nu}\rangle$ can be seen to be peaked on the lightcone ($u=0$) with width of order $\alpha$. 

Since our real interest is in the boundary theory on $\mathcal{R}\times S^3$, we need to do a conformal transformation and coordinate change to get from $\mathcal{R}^{3,1}$ to $\mathcal{R} \times S^3$. The Minkowski metric is 
\begin{equation}
ds^2 = -du dv + \frac{(v-u)^2}{4} d\Omega_2^2 .
\end{equation}
Conformally rescaling by a factor of $\frac{1}{4} (1+v^2)(1+u^2)$ and changing coordinates: $u = \tan U$, $v = \tan V$, where $U = \frac{1}{2} (\tau - \theta)$ and $V = \frac{1}{2}(\tau + \theta)$ gives the desired
\begin{equation}
ds^2 = -d \tau^2 + (d\theta^2 + \sin^2 \theta d\Omega_2^2) .
\end{equation}
Applying these transformations to the $\mathcal{R}^{3,1}$ stress tensor (\ref{eq:TPoin}) gives the $\mathcal{R}\times S^3 $ stress tensor: 

\begin{eqnarray}\label{eq:Tglobal}
\langle T_{U U} \rangle &=& \frac{2 m \alpha^4}{3 \pi^2} \frac{(1+u^2)^3 (1+v^2)}{(\alpha^2 + u^2)^3 (\alpha^2 +v^2)}\\
\langle T_{V V} \rangle &=& \frac{2 m \alpha^4}{3 \pi^2} \frac{(1+v^2)^3 (1+u^2)}{(\alpha^2 + v^2)^3 (\alpha^2 +u^2)}\\
\langle T_{U V} \rangle &=& \frac{ m \alpha^4}{3 \pi^2} \frac{(1+u^2)^2 (1+v^2)^2}{(\alpha^2 + u^2)^2 (\alpha^2 +v^2)^2} \ ,
\end{eqnarray}
where $u$ and $v$ were written above in terms of $\tau$ and $\theta$. If we are interested in the energy component, then this is given by
\begin{equation} \label{eq:T00}
\langle T_{\tau \tau} \rangle = \frac{1}{4} (\langle T_{UU}\rangle + 2 \langle T_{UV}\rangle + \langle T_{VV}\rangle).
\end{equation}

\bibliographystyle{utcaps}
\bibliography{all}

\providecommand{\href}[2]{#2}\begingroup\raggedright\begin{thebibliography}{10}

\bibitem{Bek81}
J.~D. Bekenstein, ``A universal upper bound on the entropy to energy ratio for
  bounded systems,''
{\em Phys. Rev. D} {\bf 23} (1981)  287.

\bibitem{Bek94b}
J.~D. Bekenstein, ``Do we understand black hole entropy?,''
\href{http://arxiv.org/abs/gr-qc/9409015}{{\tt gr-qc/9409015}}.

\bibitem{Tho93}
G.~'t~Hooft, ``Dimensional reduction in quantum gravity,''
\href{http://arxiv.org/abs/gr-qc/9310026}{{\tt gr-qc/9310026}}.

\bibitem{Sus95}
L.~Susskind, ``The World as a hologram,'' {\em J. Math. Phys.} {\bf 36} (1995)
  6377--6396, \href{http://arxiv.org/abs/hep-th/9409089}{{\tt hep-th/9409089}}.

\bibitem{Bou99b}
R.~Bousso, ``A covariant entropy conjecture,'' {\em JHEP} {\bf 07} (1999)  004,
\href{http://arxiv.org/abs/hep-th/9905177}{{\tt hep-th/9905177}}.

\bibitem{RMP}
R.~Bousso, ``The holographic principle,'' {\em Rev. Mod. Phys.} {\bf 74} (2002)
   825,
\href{http://arXiv.org/abs/hep-th/0203101}{{\tt hep-th/0203101}}.

\bibitem{SusWit98}
L.~Susskind and E.~Witten, ``The holographic bound in {A}nti-de~{S}itter
  space,'' \href{http://arxiv.org/abs/{h}ep-th/9805114}{{\tt
  {h}ep-th/9805114}}.

\bibitem{PolSus99}
J.~Polchinski, L.~Susskind, and N.~Toumbas, ``{Negative energy, superluminosity
  and holography},'' \href{http://dx.doi.org/10.1103/PhysRevD.60.084006}{{\em
  Phys.Rev.} {\bf D60} (1999)  084006},
\href{http://arxiv.org/abs/hep-th/9903228}{{\tt arXiv:hep-th/9903228
  [hep-th]}}.

\bibitem{HorItz99}
G.~T. Horowitz and N.~Itzhaki, ``Black holes, shock waves, and causality in the
  {AdS/CFT} correspondence,'' {\em JHEP} {\bf 02} (1999)  010,
\href{http://arXiv.org/abs/hep-th/9901012}{{\tt hep-th/9901012}}.

\bibitem{HamKab05}
A.~Hamilton, D.~N. Kabat, G.~Lifschytz, and D.~A. Lowe, ``{Local bulk operators
  in AdS/CFT: A Boundary view of horizons and locality},''
  \href{http://dx.doi.org/10.1103/PhysRevD.73.086003}{{\em Phys.Rev.} {\bf D73}
  (2006)  086003},
\href{http://arxiv.org/abs/hep-th/0506118}{{\tt arXiv:hep-th/0506118
  [hep-th]}}.

\bibitem{HamKab06}
A.~Hamilton, D.~N. Kabat, G.~Lifschytz, and D.~A. Lowe, ``{Holographic
  representation of local bulk operators},''
  \href{http://dx.doi.org/10.1103/PhysRevD.74.066009}{{\em Phys.Rev.} {\bf D74}
  (2006)  066009},
\href{http://arxiv.org/abs/hep-th/0606141}{{\tt arXiv:hep-th/0606141
  [hep-th]}}.

\end{thebibliography}\endgroup

\end{document}